\documentclass[]{spie}  

\usepackage{amsmath,amsfonts,amssymb}
\usepackage{graphicx}
\usepackage[colorlinks=true, allcolors=blue]{hyperref}

\usepackage{array}
\usepackage{url}
\usepackage{multirow}
\usepackage{silence}
\usepackage{subcaption,booktabs}

\usepackage{float}

\usepackage[misc,geometry]{ifsym} 
\usepackage{color,hyperref}

\title{CondenseUNet: A Memory-Efficient Condensely-Connected Architecture for Bi-ventricular Blood Pool and Myocardium Segmentation} 

\author[a(\Letter)]{S. M. Kamrul Hasan}
\author[a,b]{Cristian A. Linte}
\affil[a]{Center for Imaging Science, Rochester Institute of Technology, NY, USA}
\affil[b]{Biomedical Engineering, Rochester Institute of Technology, NY, USA}
\authorinfo{ Research reported in this publication was supported by the National Institute of General Medical Sciences of the National Institutes of Health under Award No. R35GM128877 and by the Office of Advanced Cyber infrastructure of the National Science Foundation under Award No. 1808530. \\
(\Letter)Further author information: \\S. M. Kamrul Hasan \texttt{(E-mail: \href{mailto:sh3190@rit.edu}{sh3190@rit.edu)}}\\  Cristian A. Linte \texttt{(E-mail: \href{mailto:calbme@rit.edu}{calbme@rit.edu})}}

\begin{document} 
\maketitle

\begin{abstract}
With the advent of Cardiac Cine Magnetic Resonance (CMR) Imaging, there has been a paradigm shift in medical technology, thanks to its capability of imaging different structures within the heart without ionizing radiation. 
However, it is very challenging to conduct pre-operative planning of minimally invasive cardiac procedures without accurate segmentation and identification of the left ventricle (LV), right ventricle (RV) blood-pool, and LV-myocardium. Manual segmentation of those structures, nevertheless, is time-consuming and often prone to error and biased outcomes. Hence, automatic and computationally efficient segmentation techniques are paramount. In this work, we propose a novel memory-efficient Convolutional Neural Network (CNN) architecture as a modification of both CondenseNet, as well as DenseNet for ventricular blood-pool segmentation by introducing a bottleneck block and an upsampling path. Our experiments show that the proposed architecture runs on the Automated Cardiac Diagnosis Challenge (ACDC) dataset using half $(50\%)$ the memory requirement of DenseNet and one-twelfth $(\sim 8\%)$ of the memory requirements of U-Net, while still maintaining excellent accuracy of cardiac segmentation. We validated the framework on the ACDC dataset featuring one healthy and four pathology groups whose heart images were acquired throughout the cardiac cycle and achieved the mean dice scores of 96.78\% (LV blood-pool), 93.46\% (RV blood-pool) and 90.1\% (LV-Myocardium). These results are promising and promote the proposed methods as a competitive tool for cardiac image segmentation and clinical parameter estimation that has the potential to provide fast and accurate results, as needed for pre-procedural planning and / or pre-operative applications.

\end{abstract}



\keywords{Cine MR image, cardiac imaging, deep learning, image segmentation, learned group convolution, ventricle blood-pool, myocardium, weight pruning}

\section{INTRODUCTION}
\label{sec:intro}
Cardiovascular diseases (CVDs) are the leading cause of death for both men and women in the United States (US) according to the American Heart Association and someone dies from distinct form of CVDs in every 38 seconds, based on 2016 data\footnote{https://newsroom.heart.org/news/nearly-half-of-all-u-s-adults-have-some-form-of-cardiovascular-disease?}. Even the number is set to reach 130 million by the year $2035$ as projected by the American Heart Association \cite{benjamin2019heart}. Nevertheless, significant scientific advances revolutionized the cardiovascular research landscape on the promise of strategical improvement to augment the diagnosis of the patients suffering from CVDs.

Cardiac Magnetic Resonance Imaging (CMRI) has made a significant paradigm shift in medical imaging through the noninvasive quantification of the volumetric changes in the heart throughout the cardiac cycle. To further study the anatomical structure and functional characteristics of the heart, dynamic short-axis cine cardiac MRI has been used extensively in recent decades. From the machine learning perspective, cardiac image segmentation is a multi-class classification problem aiming to assign each voxel, a target label. Previously, traditional machine learning techniques had been shown to achieve good performance in cardiac image segmentation \cite{peng2016review}. However, they require both prior information and manual interaction. However, manual segmentation can be susceptible to inter / intra-observer variability, hence the attempt to solve this problem automatically is desirable in the clinical setting.

The superior performance of Convolutional Neural Networks (CNNs) in solving high-level computer vision tasks has triggered the medical imaging research community to develop a powerful machine learning tool for medical image segmentation in learning intricate features in an end-to-end manner \cite{he2017mask}. 

Jain {\it et al.} \cite{jain2018efficient} designed a CNN model for cardiac image segmentation using a 2D and 3D segmentation pipeline. Isensee {\it et al.} \cite{isensee2017automatic} proposed to segment bi-ventricle and myocardium using an ensemble of modified 2D and 3D U-Net. Wolterink {\it et al.} \cite{wolterink2017automatic} designed a deep neural network for automatic cardiac segmentation, as well as disease classification from the cardiac features. Baumgartner {\it et al.} \cite{baumgartner2017exploration} explored various 2D and 3D convolution neural networks for the segmentation of the left (LV) and right (RV) ventricular cavities and the myocardium. Khened {\it et al.} \cite{khened2019fully} employed a multi-scale residual DenseNet model to automatically segment the cardiac structure from  cine MRI sequence. 
Although these methods were successful for cardiac segmentation, the use of deep model compression tasks for medical image segmentation is still rarely reported. 

\begin{figure*}[t!]

\includegraphics[width=1.0\linewidth]{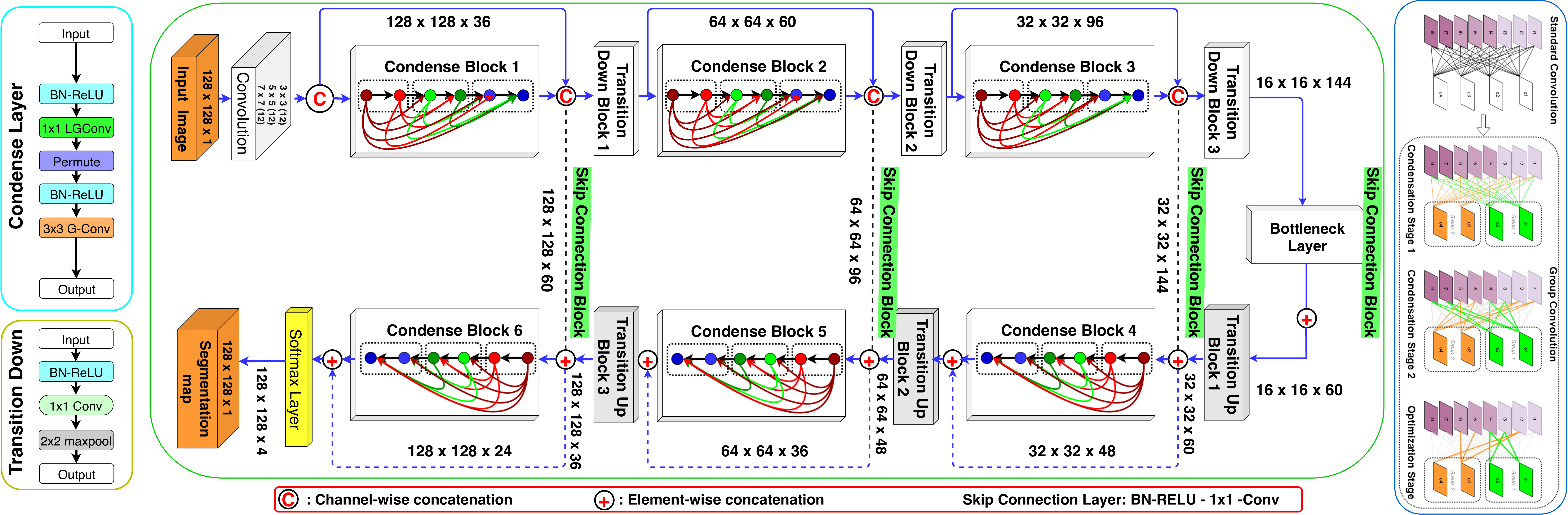}
\vspace{0.6mm}
  \caption{\textit{CondenseUNet} architecture is comprised of a downsampling path and an upsampling path. Each Condense block (CDB) consists of 3 Layers with a growth rate of k = 16. The transformations within each CDB layer are mentioned in cyan rectangular box and transition-down block in the yellow rectangular box. The concept of Learned Group Convolution (LGConv) is illustrated in the blue rectangular box and explained in Section \ref{sec:method}. The element-wise addition of Feature maps from both up and down-sampling path is matched by skip-connection block (red rectangular box). The gray blocks represent the transition-up operation which consists of $3 \times 3$ transposed convolution operation.}
\label{fig:CondenseUNet}
\end{figure*}

Although the first introduction  of group convolution in AlexNet \cite{krizhevsky2012imagenet} has been well illustrated its efficacy in recent network design, the pre-defined use of filters in each group convolution \cite{xie2017aggregated} restricts its representation capability.
Therefore, in this work, we propose to combine both learned-group convolution and weight-pruning technique with the segmentation problem in a fully convolutional setting. It is expected that \textit{CondenseUNet} will significantly improve the performance of the network in segmenting different cardiac structures (LV, RV, and LV-Myocardium) from both end-diastole and end-systole images, while also significantly reducing the number of model parameters. The proposed network architecture is shown in Figure \ref{fig:CondenseUNet}.


\section{Methodology}
\subsection{Network Architecture} \label{sec:method}
 
\textit{CondenseUNet} is both a modification of DenseNet \cite{huang2017densely}, as well as a combination of CondenseNet \cite{huang2018condensenet} and U-Net \cite{ronneberger2015u}. 
Our proposed \textit{CondenseUNet} framework substitutes the concept of both standard convolution and group convolution (G-Conv) with learned group-convolution (LG-Conv). The standard \textit{Group Convolution (GConv)} \cite{xie2017aggregated} partitions the input features into disjoint groups. Given an input tensor of shape ($I_i$, H, W), number of output channels $O_o$, and the partitioned group $M$, we apply GConv layers between input partition ($\frac{I_i}{M}$, H, W) and ($\frac{O_o}{M}$, $\frac{I_i}{M}$, $\frac{K}{h}$, $\frac{K}{w}$) weight group; where  ($\frac{K}{h}$, $\frac{K}{w}$) is the shape of the kernel, leading to the computational cost, $\mathcal{O}(\frac{I_i X O_o}{M})$. While the standard convolution needs an increased level of computation, i.e. $\mathcal{O}(I_i$ x $O_o)$, and concurrently, the pre-defined use of filters in each group convolution \cite{xie2017aggregated} restricts its representation capability, these aforementioned problems are mitigated by introducing \textit{LG-Conv} that learns group convolution dynamically during training through a multi-stage scheme, illustrated in the blue rectangular box in Figure \ref{fig:CondenseUNet}. Before training, output filters are split into equal-sized $M$ groups, where each group has its own weight. Thus, each group can select its own set of relevant input features, assisting the system to predict most relevant features at the relevant connections. This multi-stage pipeline consists of \textit{multi-condensing} stages followed by the \textit{optimization} stage.

In the first half of the pipeline, training is initiated by calculating the magnitude of the weights for each incoming feature, which are then averaged. After that, the low-magnitude weighted column is screened out from the features. Thus, a fraction of $(C -1)/C$ is truncated after each of the $C-1$ condensing stages. 

The second part of the pipeline is where all training occurs. This stage is focused on finding the optimal permutation connection that will share a similar sparsity pattern, to mitigate any negative effects on accuracy induced by the pruning process. As mentioned by Huang {\it et al.}, in their paper on CondenseNet \cite{huang2018condensenet}, both $L_1$ and $L_2$ regularizations are efficient for solving the overfitting problem, but they do not perform well as far for network optimization. To restore performance, we introduce group-lasso, an efficient regularizer that is a natural generalization of the standard lasso (least absolute shrinkage and selection operator) objective. Additionally, the group-lasso regularizer encourages group-level sparsity at the factor level by forcing all outgoing connections from a single neuron (corresponding to a group) to be either simultaneously zero or not.

As \textit{CondenseUNet} is based on both U-Net and DenseNet, it is comprised of a down-sampling and up-sampling path. The down-sampling path is similar to CondenseNet and the up-sampling path is comprised of transposed convolutions, condense blocks and skip-connections with a soft-max layer to generate the image mask. Concatenation in the skip-layer has been replaced by an element-wise addition operation to mitigate the problem of the feature-map explosion. We employ a number of layers per block as 2,3,4,5,4,3,2 with 32 initial feature maps, 3 max-pooling layers, a growth rate of k = 16, and condensation factor, C = 4 (Figure \ref{fig:CondenseUNet}).

\subsection{Cardiac MRI Data}
For this study, we used the ACDC dataset\footnote{https://www.creatis.insa-lyon.fr/Challenge/acdc/databases.html}, which is composed of short-axis cardiac cine-MR images acquired from 100 different patients divided into 5 evenly distributed subgroups according to their cardiac condition: normal- NOR, myocardial infarction- MINF, dilated cardiomyopathy- DCM, hypertrophic cardiomyopathy- HCM, and abnormal right ventricle- ARV, available as a part of the STACOM 2017 ACDC challenge\cite{bernard2018deep}. 




\subsection{Data Pre-processing}
To solve the class-imbalance problem in multi-slice cardiac MR images, a patch of size $128 \times 128$ was extracted around the LV center from a full-sized cardiac MR and slice-wise normalization of voxel intensities were performed. The training dataset was divided into 70\% training data, 15\% validation data, and 15\% testing data with five non-overlapping folds for cross-validation. We heavily augment the ACDC dataset through both affine and elastic transformations, including several operations: (i) re-scaling: random zoom factor ranging $0.8 \sim 1.2$, (ii) translation: random shift ranging $-5\sim 5mm$, (iii) rotation, and (iv) Gaussian noise addition. 



\subsection{Loss Function and Evaluation Metrics}

We use a dual-loss function that incorporates both Cross-Entropy and Dice loss which is formulated in the equation \ref{eq:1} and also illustrated in our paper, \cite{8856791}U-NetPlus:

\begin{equation} \label{eq:1}
\mathcal{L}_{Dual-Loss} = \alpha.\mathcal{L}_{Entropy}(x, y_1|\theta) + \beta. (1-\mathcal{L}_{Dice}(x, y_1|\theta))
\end{equation}
\noindent
where $\mathcal{L}_{Entropy}$ is the cross-entropy loss and $\mathcal{L}_{Dice}$ is the dice loss. The symbol, $\theta$ denotes the network weights, $x$ denotes the training samples, $y_{1}$ be the label map of the training set, $\alpha$ and $\beta$ denotes the weighting parameters. Additionally, we evaluate several clinical indices, including myocardial mass and ejection fraction\cite{wagholikar2018extraction}. All assessment metrics were evaluated and compared between our proposed method and the ground truth.

\subsection{Network Training and Testing}
The Networks implemented in Tensorflow\footnote{https://www.tensorflow.org/} were initialized with He normal initializer \cite{he2015delving} and trained for 200 epochs with a batch size of 16. We used the Adam optimizer with a learning rate of 0.0001. The standard DenseNet has a growth rate of k = 12; in our case, k was 16. All experiments were conducted on a machine equipped with an NVIDIA RTX 2080 Ti GPU (11GB memory).

\begin{figure*}[t!]
\includegraphics[width=1.0\linewidth]{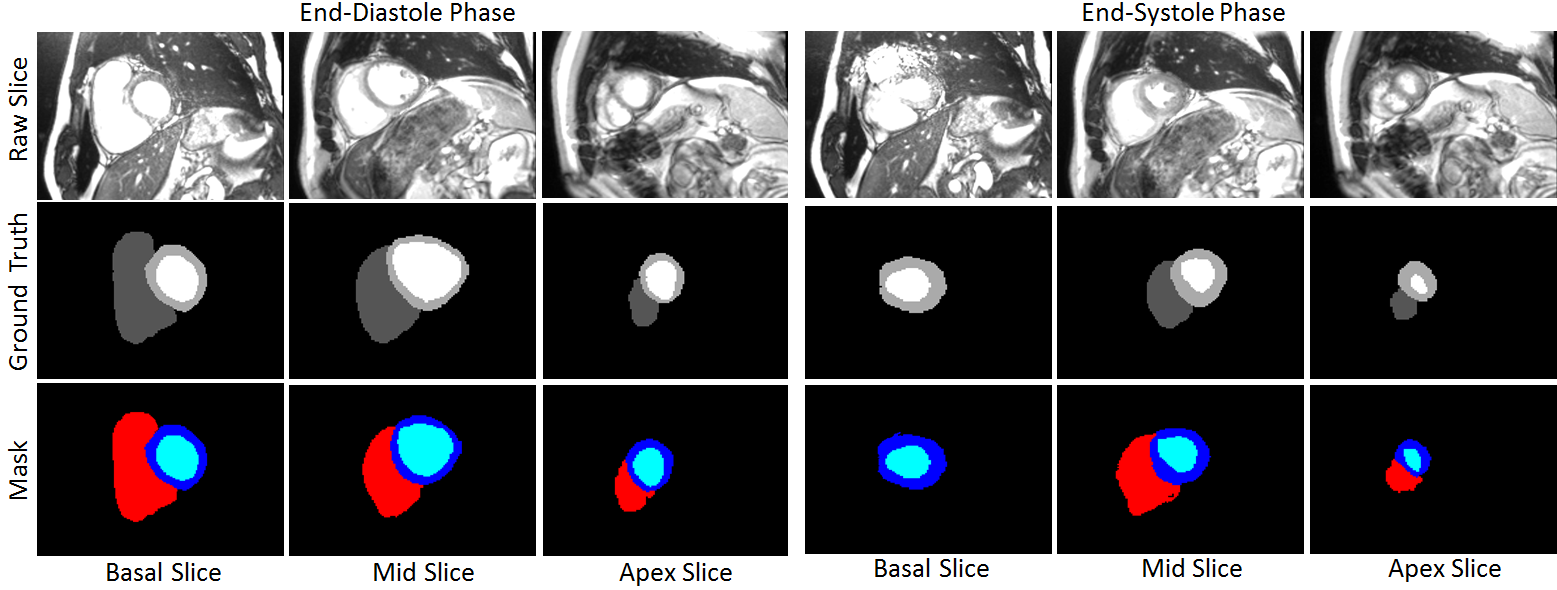}
  \caption{Segmentation results at both ED and ES phases from the base (high slice index) to apex (low slice index) showing RV in red, LV in cyan, and Myo in blue.}
\label{fig:results}
\end{figure*}



\scriptsize
\begin{table*}[t!]
{\caption{Comparison of the segmentation results: Dice scores (\%) on ACDC dataset for RV blood-pool, LV-myocardium, and LV blood-pool, obtained from all best networks against proposed model \textbf{CondenseUNet: CUNet}. UNet by [Jain {\it et al.}]  \cite{patravali20172d}, MUNet: Modified 3D UNet by [Baumgartner {\it et al.}]\cite{baumgartner2017exploration}, MNet: Modified M-Net by [Jang {\it et al.}]\cite{jang2017automatic}, and DNet: DenseNet by [Khened {\it et al.}]\cite{khened2019fully}.}
\vspace{1.4mm}
\label{tab:SegmentationACDC}}
\begin{subtable}[t]{\textwidth}
\caption{Right Ventricle (RV) Blood-pool Segmentation Evaluation}
\vspace{-3mm}
\label{tab:RVSegmentation}
\begin{center}
\begin{tabular}{|p{0.65cm}||>{\centering\arraybackslash}m{0.9cm}>{\centering\arraybackslash}m{0.9cm}|>{\centering\arraybackslash}m{0.9cm}>{\centering\arraybackslash}m{0.9cm}|>{\centering\arraybackslash}m{0.9cm}>{\centering\arraybackslash}m{0.9cm}||>{\centering\arraybackslash}m{0.9cm}>{\centering\arraybackslash}m{0.9cm}|>{\centering\arraybackslash}m{0.9cm}>{\centering\arraybackslash}m{0.9cm}|>{\centering\arraybackslash}m{0.9cm}>{\centering\arraybackslash}m{0.85cm}|}

\hline
\multirow{2}{1.4cm} & \multicolumn{6}{c||}{End Diastole (ED)} & \multicolumn{6}{|c|}{End Systole (ES)}\\ 
\cline{2-13}
& UNet & DCN & MUNet & MNet & DNet & CUNet & UNet & DCN & MUNet & MNet & DNet & CUNet \\
\hline \hline
Dice & 91.0 & 92.0 & 93.20 & 92.90 & {\bf 93.50} & 93.46  & 83.0 & 84.0 & 88.30  & 88.50 &  {\bf87.90} & 87.60 \\


\hline


\end{tabular}
\end{center}
\end{subtable}

\vspace{2mm}

\begin{subtable}[t]{\textwidth}
\caption{LV-Myocardium Segmentation Evaluation}
\vspace{-3mm}
\label{tab:LVSegmentation}
\begin{center}
\begin{tabular}{|p{0.63cm}||>{\centering\arraybackslash}m{0.9cm}>{\centering\arraybackslash}m{0.9cm}|>{\centering\arraybackslash}m{0.9cm}>{\centering\arraybackslash}m{0.9cm}|>{\centering\arraybackslash}m{0.9cm}>{\centering\arraybackslash}m{0.9cm}||>{\centering\arraybackslash}m{0.9cm}>{\centering\arraybackslash}m{0.9cm}|>{\centering\arraybackslash}m{0.9cm}>{\centering\arraybackslash}m{0.9cm}|>{\centering\arraybackslash}m{0.9cm}>{\centering\arraybackslash}m{0.9cm}|}

\hline
\multirow{2}{1.4cm} & \multicolumn{6}{c||}{End Diastole (ED)} & \multicolumn{6}{|c|}{End Systole (ES)}\\ 
\cline{2-13}
& UNet & DCN & MUNet & MNet & DNet & CUNet & UNet & DCN & MUNet & MNet & DNet & CUNet \\
\hline \hline
Dice & 86.0  & 86.0 &  89.2  & 87.0  & 88.9 & {\bf 89.43}  & 88.0  & 88.0  & 90.1  & 88.8 & 89.8  & {\bf 90.1} \\
\hline


\end{tabular}
\end{center}
\end{subtable}

\vspace{2mm}

\begin{subtable}[t]{\textwidth}
\caption{Left Ventricle (LV) Blood-pool Segmentation Evaluation}
\vspace{-3mm}
\label{tab:SegmentationLV}
\begin{center}
\begin{tabular}{|p{0.6cm}||>{\centering\arraybackslash}m{0.9cm}>{\centering\arraybackslash}m{0.9cm}|>{\centering\arraybackslash}m{0.9cm}>{\centering\arraybackslash}m{0.9cm}|>{\centering\arraybackslash}m{0.9cm}>{\centering\arraybackslash}m{0.9cm}||>{\centering\arraybackslash}m{0.9cm}>{\centering\arraybackslash}m{0.9cm}|>{\centering\arraybackslash}m{0.9cm}>{\centering\arraybackslash}m{0.9cm}|>{\centering\arraybackslash}m{0.9cm}>{\centering\arraybackslash}m{0.9cm}|}

\hline
\multirow{2}{1.4cm} & \multicolumn{6}{c||}{End Diastole (ED)} & \multicolumn{6}{|c|}{End Systole (ES)}\\ 
\cline{2-13}
& UNet & DCN & MUNet & MNet & DNet & CUNet & UNet & DCN & MUNet & MNet & DNet & CUNet \\
\hline \hline
Dice & 95.0  & 96.0  & 96.3  & 96.1 & 96.4 & {\bf96.78} & 90.0 & 91.0 & 91.1 & 91.5  & 91.7  & {\bf 95.1} \\
\hline



\end{tabular}
\end{center}
\end{subtable}

\end{table*}
\normalsize



\begin{figure}[t!]
\centering
\includegraphics[width=0.8\linewidth]{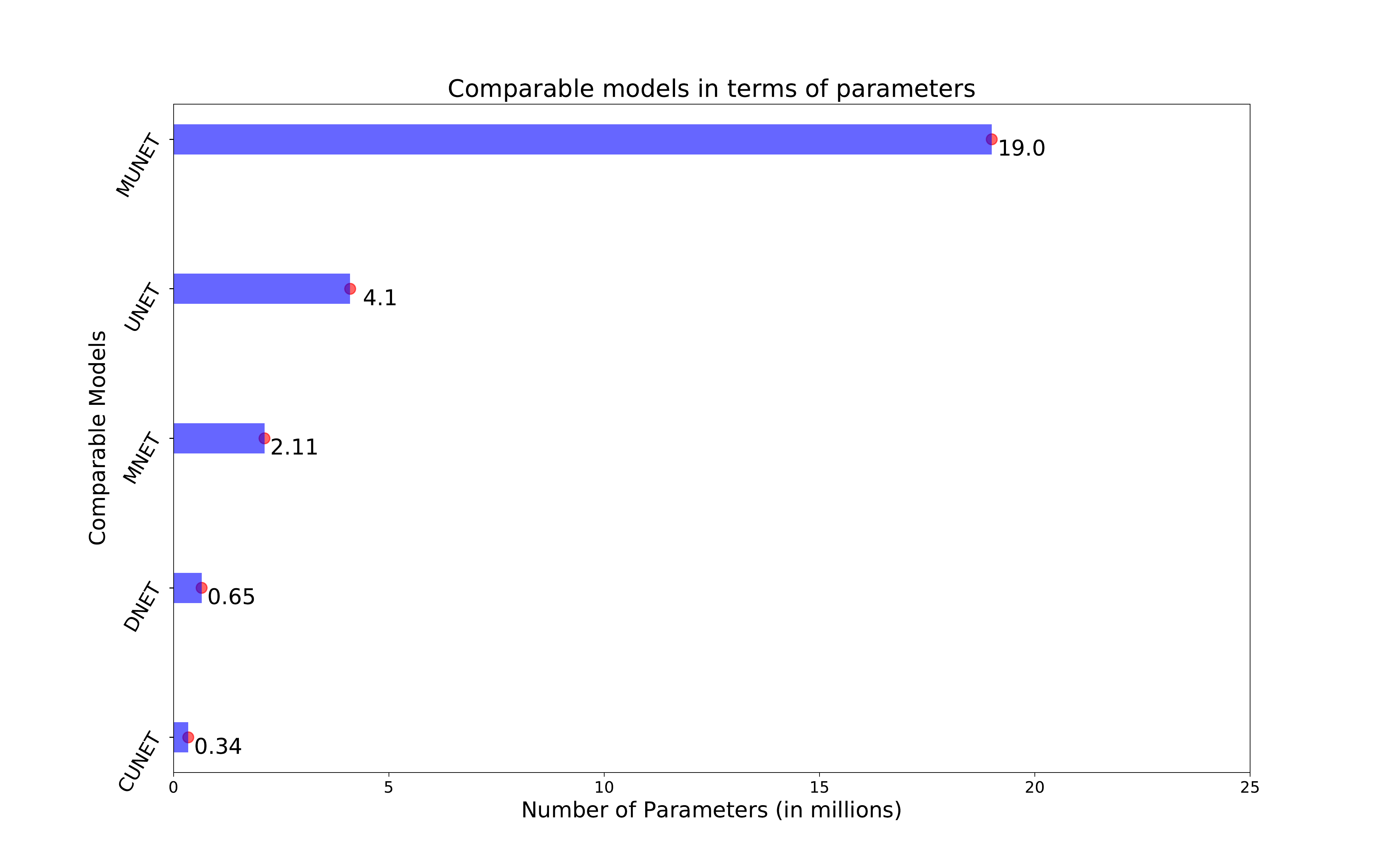}
\vspace{2mm}
  \caption{ Comparison of \textit{CondenseUNet} with other convolutional networks on the ACDC dataset in terms of number of model parameters ($\times 10^6$) -- MUNet: Modified 3D UNet, MNet: Modified M-Net, DNet: DenseNet, CUNet: CondenseUNet.}
\label{fig:param}
\end{figure}

\section{ Results}
In this section we evaluated the performance of our proposed \textit{CondenseUNet} in terms of geometric scores, and number of parameters. Table \ref{tab:SegmentationACDC} summarizes the Dice score, which measure the accuracy of the segmentation of the left and right ventricle blood-pool in diastole and systole and the left ventricle myocardium vis-a-vis their corresponding ground truth segmentation. The table also shows that the segmentation accuracy is generally better than the best comparable architecture for both LV and LV-myocardium in both ED and ES phase. An illustration of the qualitative segmentation results in both end-diastole (ED) and end-systole (ES) is provided in Figure \ref{fig:results}. As shown, the segmentation mask for the ventricles and LV-myocardium performs best in both end-diastole and end-systole phase. Nevertheless, with the exception of several instance of the RV blood-pool segmentation, \textit{CondenseuNet} yields the highest segmentation performance, on average, and significantly surpasses all other network architectures in terms of the number of model parameters.

Figure \ref{fig:param} plots the number of parameters (in millions) for all the comparable architectures. It shows that our proposed architecture (CUNet) runs on the ACDC dataset using half $(50\%)$ the memory requirement of DenseNet and one-twelfth $(\sim 8\%)$ of the memory requirements of U-Net, while still achieving comparable segmentation performance.  
\vspace{-2mm}

\section{Conclusion and Future Work}
In this paper, we propose a new paradigm for accurate LV, RV blood pool and myocardium segmentation from cine cardiac MR images by combining the memory-efficient CondenseNet architecture with the modified U-Net model. The capability of our network to learn the group structure allows multiple groups to re-use the same features via dense connectivity. Moreover, the integration of efficient weight pruning with a simple regularizer leads to high computational savings without compromising the accuracy of the segmentation and the fidelity of the estimated clinical parameters. The computationally efficient and accurate segmentation masks obtained from our proposed method can be used for pre-operative generation of subject-specific LV, RV and LV-myocardium models for surgical planning, navigation, and guidance applications.

Our proposed work reveals that a properly designed condensely connected network, when trained in the U-Net shaped framework, produces significantly higher performance with fewer trainable parameters. According to DICE metrics, we are achieving 96.78\% DICE for LV blood pool in ED and 95.1\% in ES phase, which showed at least 0.4\% improvement in ED phase and 3.7\% improvement in ES phase over the current methods, as well as more than 5\% improvement over the standard U-Net architecture. We observed that the segmentation results for RV have not improved significantly beyond those of the LV or myocardium. 

An alternative solution for better segmentation of the RV would be to perform architecture-agnostic efficient inference with cosine learning rate, as well as slice refinement. For future work, we will explore an additional attention analysis to further analyze the segmentation performance. This study will help us understand and visualize where our algorithm ``looks'' in an image by using a novel image saliency technique. Therefore, this computationally efficient segmentation model can be used for the generation of subject-specific models of the cardiac anatomy for computer-aided diagnosis or minimally invasive therapy planning.

\acknowledgments 
Research reported in this publication was supported by the National Institute of General Medical Sciences of the National Institutes of Health under Award No. R35GM128877 and by the Office of Advanced Cyber infrastructure of the
National Science Foundation under Award No. 1808530.

\bibliography{main.bib} 
\bibliographystyle{spiebib} 

\mbox{}
\vfill
© Preprint submitted to ArXiv by S M Kamrul Hasan
\end{document}